\documentclass[pre,twocolumn,showpacs]{revtex4}
\usepackage{dcolumn}
\usepackage{amsmath}
\usepackage{amssymb}
\usepackage{amsbsy}
\usepackage{graphicx}

\begin{document}
\title{Coefficient of Restitution for Viscoelastic Spheres: The Effect of Delayed Recovery}

\author{Thomas Schwager and Thorsten P\"{o}schel}  
\affiliation{Charit\'e, Augustenburger Platz 1, 13353 Berlin, Germany}

\date{\today}

\begin{abstract}
The coefficient of normal restitution of colliding viscoelastic spheres is computed as a function of the material properties and the impact velocity. From simple arguments it becomes clear that in a collision of purely repulsively interacting particles, the particles loose contact slightly before the distance of the centers of the spheres reaches the sum of the radii, that is, the particles recover their shape only after they lose contact with their collision partner. This effect was neglected in earlier calculations which leads erroneously to attractive forces and, thus, to an underestimation of the coefficient of restitution. As a result we find a novel dependence of the coefficient of restitution on the impact rate.
\end{abstract}


\maketitle

\section{Introduction}

The dynamics of a granular system is governed by the particle interaction law, that is, by the forces the particles in contact exert on one another. In general, these forces may be complicated functions of the time dependent mutual deformation and relative velocities in normal and tangential direction. In the simplest case the particles are modeled as spheres interacting via normal and tangential forces. 

Given particles of radii $R_{1/2}$ and masses $m_{1/2}$ at positions $\vec{r}_1(t)$ and $\vec{r}_2(t)$ traveling at velocities $\vec{v}_1(t)$ and $\vec{v}_2(t)$. The particle deformation is then described by
\begin{equation}
\xi(t)\equiv\max\left(0,2R-\left|\vec{r}_1-\vec{r}_2\right|\right)
\end{equation}
and the deformation rate $\dot{\xi}(t)$. Apart from material properties,
the dissipative and elastic
components of the normal force of particles in contact depend on the
deformation, the deformation rate, and the radii,
\begin{equation}
  F=F^\text{(el)}\left(\xi, R_1, R_2\right) +  F^\text{(dis)}\left(\xi,\dot{\xi}, R_1, R_2\right)\,.
\label{eq:F}
\end{equation}
The functional form of these forces is model specific, see e.g.
\cite{SchaeferDippelWolf:1995,KruggelEtAl:2007,algo}. Having specified the interaction
forces, the dynamics of a ensemble of granular particles can be solved
by a (force-based) Molecular Dynamics scheme.

An alternative approach uses the concept of the coefficient of
restitution, relating the normal component of a pair of particles
before and after a collision,
\begin{equation}
  \varepsilon\equiv -\left.\dot{\xi}^\prime\right/\dot{\xi}\,.
  \label{eq:eps1}
\end{equation}
This concept does not consider the duration of a contact, that is, a collision is an instantaneous event. Consequently, it is assumed that the particles collide exclusively pairwise. This condition is justified if the mean flight time between collisions is much larger than the duration of a collision which restricts the range of applicability of the coefficient of restitution. The material properties of the particles are, thus, assumed to assure short duration of contact and/or the particle number density of the system should be small enough (low collision frequency) to neglect multi-particle contacts. In practical applications, event-driven Molecular Dynamics simulations, based on the coefficient of restitution, deliver frequently satisfying results even for rather dense systems.

Both concepts, interaction forces and the coefficient of restitution, can be applied to describe the dynamics of a granular system using either (force-based) Molecular Dynamics or event-driven Molecular Dynamics. Describing the same physical systems, of course, the coefficient of restitution and the interaction forces must be closely related. Indeed, integrating Newton's equation of motion for an isolated pair of particles colliding at time $t=0$,
\begin{equation}
m_\text{eff} \ddot{\xi}+F\left(\dot{\xi},\xi\right)=0\,;~~~~\xi(0)=0\;;~~~~ \dot{\xi}=v
\label{eq:EOM}
\end{equation}
with $m_\text{eff}\equiv m_1m_2/\left(m_1+m_2\right)$ and 
\begin{equation}
v\equiv \frac{\left[\vec{v}_1(0)-\vec{v}_2(0)\right]\cdot \left[\vec{r}_1(0)-\vec{r}_2(0)\right]}{R_1+R_2}
\label{eq:vdef}
\end{equation}
to obtain the trajectory $\xi(t)$, the coefficient of restitution is
\begin{equation}
\varepsilon= -\left.\dot{\xi}(t_c)\right/v\,,
\label{eq:COR1}
\end{equation}
where $t_c$ is the duration of the collision. This computation was performed for several interaction force models \cite{SchaeferDippelWolf:1995,SchwagerPoeschel:1998,RamirezEtAl:1999,KruggelEtAl:2007}. Albeit conceptually simple, even for simple force laws the algebra is rather technical.

It is important that Eq. \eqref{eq:F} applies to {\em particles in contact}. Obviously, in the absence of adhesion, the interaction force between colliding granular particles is strictly repulsive. Formally, however, during the decompression phase where $\dot{\xi}<0$ the dissipative term $F^\text{(dis)}$ in Eq. \eqref{eq:F} may overcompensate the pure repulsive conservative force $F^\text{(el)}$ erroneously yielding an attractive total force, e.g. \cite{Luding98,LudingHabil}. 

In Molecular Dynamics simulations, therefore, the normal force between particles is usually computed as $F^*=\max(0,F)$, with $F$ given in Eq. \eqref{eq:F} which assures that only repulsive forces act. The force $F^*$ can, thus, be conveniently used in simulations.

The described artifact of negative interaction force originates from an inappropriate definition of the end of a collision at time $t_c$. The duration of the collision, $t_c$, however, is needed for the derivation of the coefficient of restitution by means of Eq. \eqref{eq:COR1}. Whereas the beginning of a contact is well described by the condition $\xi(0)=0$, the end of a collision at time $t_c$ is less trivial.  

For simplicity of the computation in the literature it was assumed that the end of a collision is determined by the condition
\begin{equation}
  \label{eq:oldbc}
  \xi(t_c)=0~~~\text{with}~~~t_c>0\,.  
\end{equation}
As described above, in the decompression phase it may happen that $F\left(\xi,\dot{\xi}\right)<0$. This means the collision may be completed even before $\xi=0$. Thus, the surfaces of the particles lose contact slightly before the distance of their centers exceeds the sum of their radii. Consequently, the deformation of the particles may last longer than the time of contact and the particles gradually recover their spherical shape {\em after} they lost contact. The definition of the end of a collision
\begin{equation}
  \label{eq:newbc}
  F(t_c)=0~~~\text{with}~~~t_c>0
\end{equation}
takes the described scenario into account and assures that the particles interact exclusively repulsively.

Obviously since erroneous attractive forces are excluded by the improved condition for the end of collision, the resulting coefficient of restitution is expected to be larger for the definition Eq. \eqref{eq:newbc} than the value obtained for the condition Eq. \eqref{eq:oldbc}.

Let us demonstrate the influence of the definition of $t_c$ to the coefficient of restitution for the simplest form of the interaction force, the linear dash-pot
\begin{equation}
  \label{eq:dashpot}
  F\left(\xi,\dot{\xi}\right)=-k\xi - \gamma\dot{\xi}\,.
\end{equation}
Although neither the elastic nor the dissipative components are appropriate for the description of dissipatively colliding spheres (see below), the linear dash-pot model is frequently used in Molecular Dynamics simulations of granular systems. The main advantage of this model is the impact-velocity independent coefficient of restitution which follows from Eq. \eqref{eq:COR1}. Using the condition \eqref{eq:oldbc}, we obtain for the case of low damping (e.g. \cite{SchaeferDippelWolf:1995,KruggelEtAl:2007})  
\begin{equation}
  \label{eq:eps_dashpot_old}
  \varepsilon=\exp\left(-\frac{\beta\pi}{\omega}\right)~~~\text{and}~~~t_c=\frac{\pi}{\omega}
\end{equation}
with $\omega\equiv\sqrt{\omega_0^2-\beta^2}$; $\omega_0\equiv \sqrt{k/m_\text{eff}}$; $\beta\equiv \gamma/2m_\text{eff}$. 
Obviously, this result contradicts the assumption of non-attractive interaction since 
\begin{equation}
F\left(t_c\right) = F\left(\xi\left(t_c\right),\dot{\xi}\left(t_c\right)\right)=F\left(0,-\varepsilon v\right) = \gamma\varepsilon v > 0\,.
\end{equation}

For the condition Eq. \eqref{eq:newbc} for $t_c$, taking into account that there is only repulsive interaction between granular particles we find \cite{SchwagerPoeschel:2007}
\begin{equation}
  \varepsilon_n = \begin{cases}
    \displaystyle\exp\left[-\frac{\beta}{\omega_n}\left(\pi - \arctan\displaystyle\frac{2\beta\omega_n}{\omega_n^2-\beta^2}\right)\right],~~~  \displaystyle\beta<\frac{\omega_0}{\sqrt{2}} \\[0.3cm]
    \displaystyle\exp\left[-\frac{\beta}{\omega_n}\arctan\displaystyle\frac{2\beta\omega_n}{\omega_n^2-\beta^2}\right],
    ~~~ \displaystyle \beta\in\left[\frac{\omega_0}{\sqrt{2}},\omega_0\right]
\\[0.3cm]
    \displaystyle\exp\left[-\frac{\beta}{\omega_n}\ln\frac{\beta+\omega_n}{\beta-\omega_n}\right],~~~ \beta>\omega_0
    \end{cases}
  \label{eq:dashpotCOR1}
\end{equation}
It can be shown that the solutions, Eq. \eqref{eq:eps_dashpot_old} and \eqref{eq:dashpotCOR1} are fundamentally different: for values of the parameter $\beta/\omega_0$ above one the duration of the collision $t_c$ diverges in case of Eq. \eqref{eq:eps_dashpot_old}, that is, $\varepsilon=0$. Thus, the particles collide with finite velocity and stick together (dissipative capture), despite our precondition of purely non-attractive interaction. The solution Eq. \eqref{eq:dashpotCOR1} does not reveal this unphysical behavior. For a detailed discussion see \cite{SchwagerPoeschel:2007}.

The linear dash-pot model serves here only as an example to show that even for the simplest force laws the adequate characterization of the end of the collision modifies the known results for the coefficient of restitution in a non-trivial way. For the case of the linear dash-pot, the definition of $t_c$, Eqs. \eqref{eq:oldbc} or \eqref{eq:newbc}, changes the coefficient of restitution as a function of the material parameters $k$ and $\gamma$ however, $\varepsilon$ is independent of the impact velocity $v$ in both cases. 

It is the aim of this paper to compute the coefficient of restitution for the simplest physically consistent force law for viscoelastic spheres with regard to the definition Eq. \eqref{eq:newbc} for $t_c$. We will see that the appropriate choice of the condition for the end of the collision does not only change the dependence of the coefficient of restitution on the material parameters but also the functional form of its dependence on the impact velocity.

As our main result we will show that for the definition of $t_c$ given by Eq. \eqref{eq:newbc} the coefficient of restitution $\varepsilon$ is given by a series in powers of $v^{1/10}$ whereas for the definition of $t_c$ according to Eq. \eqref{eq:oldbc} $\varepsilon$ is a series in powers of $v^{1/5}$ \cite{SchwagerPoeschel:1998}.

\section{Viscoelastic spheres}

We write the interaction force law for viscoelastic spheres \cite{BrilliantovEtAl:1996} as
\begin{equation}
  \label{eq:ViscoForce}
  F\left(\xi,\dot{\xi}\right) =  -\rho \xi^{3/2}-\frac32 A\rho\sqrt{\xi}\dot{\xi}\,.
\end{equation}
The elastic part is given by the Hertz contact force \cite{Hertz:1882} with the elastic constant
\begin{equation}
  \label{eq:rhodef}
  \rho\equiv \frac{2Y\sqrt{R_\text{eff}}}{3\left(1-\nu^2\right)}
\end{equation}
where $Y$ is the Young modulus, $\nu$ is the Poisson ratio and the effective radius of the colliding pair $R_\text{eff}\equiv R_i R_j /\left(R_i+R_j\right)$. The dissipative part, $\sim\sqrt{\xi}\dot{\xi}$, was derived independently in \cite{KuwabaraKono:1987,BrilliantovEtAl:1996,MorgadoOppenheim:1997} using different methods but only the method in \cite{BrilliantovEtAl:1996} allows to derive the dissipative constant
\begin{equation}
  \label{eq:Adef}
  A\equiv \frac13\frac{\left(3\eta_2-\eta_1\right)^2}{3\eta_2+2\eta_1}\left[\frac{\left(1-\nu^2\right)\left(1-2\nu\right)}{Y\nu^2}\right]
\end{equation}
as a function of  viscous material constants $\eta_{1/2}$ that relate the dissipative stress tensor to the deformation rate tensor \cite{Landau} and the elastic constants $Y$ and $\nu$.

While the coefficient of restitution for the linear dash-pot model depends only on the material constants, it may be shown already from a dimension analysis that for viscoelastic particles the coefficient of restitution cannot be independent of the impact velocity, \cite{Tanaka:1991,TsujiTanakaIshida:1991,Taguchi:1992JDP, LudingClementBlumenRajchenbachDuran:1994DISS,RamirezEtAl:1999}. It may be shown, moreover, either by scaling arguments \cite{RamirezEtAl:1999} or in a more accurate way by a rather technical analysis \cite{SchwagerPoeschel:1998} that the coefficient of restitution depends on the impact velocity as $\varepsilon=\varepsilon\left(v^{1/5}\right)$. The coefficient of restitution was obtained in \cite{SchwagerPoeschel:1998} for the definition \eqref{eq:oldbc} as a series expansion in powers of $v^{1/5}$. (For an equivalent derivation for viscoelastic discs see \cite{Schwager}.) In the following we derive the coefficient if restitution for the end of the collision given by Eq. \eqref{eq:newbc}.

\section{Equation of Motion}

Newton's equation of motion for the collision of viscoelastic spheres reads
\begin{equation}
  \label{eq:viscoNewton}
  \ddot{\xi} + k\xi^{3/2} + \gamma \sqrt{\xi}\dot{\xi} = 0
\end{equation}
with initial conditions
\begin{equation}
  \label{eq:viscoInitial}
  \xi(0)=0\,;~~~~~\dot{\xi}=v\,.
\end{equation}
and the constants 
\begin{equation}
  \label{eq:viscoConstants}
  k\equiv \frac{\rho}{m_\text{eff}}\,;~~~~~\gamma=\frac32\frac{\rho A}{m_\text{eff}}\,.
\end{equation}
The natural unit of time is $t_\text{scale}=k^{-2/5}v^{-1/5}$ which is proportional to the duration of the undamped collision and the natural unit of length is $\xi_\text{scale}=k^{-2/5}v^{4/5}$ which is proportional to the maximal deformation. Adopting both natural units would reduce the number of free parameters to one which reads $\gamma k^{-3/5}v^{1/5}$ \cite{RamirezEtAl:1999}. This indicates that the coefficient of restitution is a function of $v^{1/5}$. For reasons which will become clear in the course of the computation (see explanation at Eq. \eqref{eq:equalxmax}) it is not advisable to use the natural unit of length. Instead we adopt the length scale $\xi_\text{scale}=k^{-2/5}$. We, thus, scale time and length as: 
\begin{equation}
  \xi = \frac{x}{k^{2/5}}\,;~~~~ t = \frac{\tau}{k^{2/5}v^{1/5}}
\label{eq:scales}
\end{equation}
and arrive at the equation
\begin{equation}
  \label{eq:trajectory}
  \begin{split}
  &\ddot{x} + \beta v^{-1/5}\dot{x}\sqrt{x} + v^{-2/5}x^{3/2} = 0\\
  &x(0)= 0\,;~~~\dot{x}(0) = v^{4/5}\,,
  \end{split}
\end{equation}
where dots mean derivatives with respect to the scaled time $\tau$ and $\beta\equiv\gamma k^{-3/5}$. Note that the deformation $\xi$ or $x$ are counted positive if the particles deform each other. The impact velocity $\dot{\xi}(0)$ or $\dot{x}(0)$ has to be positive as its action {\em increases} the deformation. 

\section{Trajectory}

First we have to determine the trajectory of the particles during the collision. To this end we apply the method which was introduced in \cite{SchwagerPoeschel:1998}. 

First we observe that the trajectory cannot be a series in integer powers of time due to the fact that the third and higher time derivatives of the deformation are singular at $x=0$. The deformation $x=0$ corresponds to the start of the collision and also to its end under the condition Eq. \eqref{eq:oldbc}. (Here we consider the collision for the condition Eq. \eqref{eq:newbc}, nevertheless, for the calculation we refer in several places to the end of the collision due to Eq. \eqref{eq:oldbc} which we call the {\em na\"{\i}ve} end of the collision.) As an example for such a divergence, the third time derivative of $x$ reads:
\begin{equation}
  x^{(3)} = -\frac{3}{2v^{2/5}}\dot{x}\sqrt{x} + \frac{\beta}{v^{3/5}} x^2 + \frac{\beta^2}{v^{2/5}}x\dot{x} - \frac{\beta}{2v^{1/5}}\frac{\dot{x}^2}{\sqrt{x}}
\end{equation}
The last term diverges for $x\to0$ as for the beginning and the end of collision $\dot{x}\ne0$. It turns out that instead of integer powers the trajectory is a series of half-integer powers of $\tau$. The computation of the trajectory $x(\tau)$ is explained in detail in appendix \ref{sec:appendix:trajectory}. The first few terms read
\begin{equation}
  \begin{split}
\label{eq:trajectory2}
  x &= v^{4/5}\left[\tau-\frac{4}{15}\beta v^{1/5}\tau^{5/2} - \frac{4}{35}\tau^{7/2} + \frac{1}{15}\beta^2v^{2/5}\tau^4\right.\\
  & \left. + \frac{3}{70}\beta v^{1/5}\tau^5 - \frac{38}{2475}\beta^3v^{3/5}\tau^{11/2} + \frac{1}{175}\tau^6\right.\\
  & \left. - \frac{937}{75075}\beta^2v^{2/5}\tau^{13/2} + \frac{2612}{779625}\beta^4v^{4/5}\tau^7\right.\\
  &\left. - \frac{713}{238875}\beta v^{1/5}\tau^{15/2} + \frac{43943}{13513500}\beta^3v^{3/5}\tau^8\right.\\
  &\left. -\left(\frac{22}{104125}+\frac{31159}{44178750}\beta^5v\right)\tau^{17/2}\right.\\
  & \left. + \frac{871}{808500}\beta^2v^{2/5}\tau^9\right.\\
  & \left. -\frac{192113}{242492250}\beta^4v^{4/5}\tau^{19/2}\right] + {\cal O}\left(\tau^{10}\right)
\end{split}
\end{equation}
It turns out that this series converges very slowly which means that we need the series up to a high order (see below). The structure of this result becomes clear if we sort the terms in escalating powers of the damping parameter $\beta$. The trajectory then takes the form
\begin{equation}
  \begin{split}
  x &= v^{4/5}\left(\tau - \frac{4}{35}\tau^{7/2} + \frac{1}{175}\tau^6 - \frac{22}{104125}\tau^{17/2} + \ldots\right)\\
  & + \beta v\left(- \frac{4}{15}\tau^{5/2} + \frac{3}{70}\tau^5 - \frac{713}{238875}\tau^{15/2} + \ldots\right)\\
  & + \beta^2v^{6/5}\left(\frac{1}{15}\tau^4 - \frac{937}{75075}\tau^{13/2} + \frac{871}{808500}\tau^9 + \ldots\right)\\
  & + \beta^3v^{7/5}\left( - \frac{38}{2475}\tau^{11/2} + \frac{43943}{13513500}\tau^8 + \ldots\right)\\
  & + \beta^4v^{8/5}\left(\frac{2612}{779625}\tau^7 -\frac{192113}{242492250}\tau^{19/2}\right)\\
  & + \beta^5v^{9/5}\left(- \frac{31159}{44178750}\tau^{17/2} + \ldots\right) + {\cal O}\left(\tau^{10}\right)
  \end{split}
  \label{eq:trajectory1}
\end{equation}
The expressions in brackets do not contain any parameter except for pure numbers. They are, hence, universal functions which we shall call $x_i(\tau)$, where the index $i$ gives the power of $\beta$ it is associated with. Note furthermore that subsequent powers of $\tau$ in each function differ by $5/2$. The trajectory can be written compactly as
\begin{multline}
  x(\tau) = v^{4/5}x_0(\tau) + \beta v x_1(\tau) + \beta^2v^{6/5}x_2(\tau) +\ldots\\
  = v^{4/5}\sum_{k=0}^\infty\left(\beta v^{1/5}\right)^kx_k(\tau)
\end{multline}
The function $x_0$ is the trajectory of the undamped ($\beta=0$) collision. It is known \cite{RamirezEtAl:1999} that it reaches its maximal compression at time
\begin{equation}
  \tau_\text{max}^0 = \left(\frac{4}{5}\right)^{3/5}\frac{\Gamma\left(\displaystyle\frac{2}{5}\right)\Gamma\left(\displaystyle\frac{1}{2}\right)}{2\Gamma\left(\displaystyle\frac{9}{10}\right)} \approx 1.609 \,.
\label{eq:tchhalf}
\end{equation}
The total duration of the undamped collision is $\tau_c^0=2\tau_\text{max}^0$ as the undamped trajectory is symmetrical with respect to the point of maximal compression. 

We proceed with computing the time of maximal compression of the damped problem along with the value of maximal compression. We use the Ansatz
\begin{equation}
  \tau_\text{max} = \tau_{\rm max}^0 + \sum_{k=1}^{\infty}a_k\beta^kv^{k/5}
\end{equation}
and solve for the coefficients $a_k$ as explained in detail in Appendix \ref{sec:appendix:maxcomp}. The first coefficients $a_k$  are
listed in Table \ref{tab:coefficients}. The principal form of these and other similar expressions -- power series in $\beta v^{1/5}$ -- can be derived by scaling arguments detailed in \cite{RamirezEtAl:1999}. The maximal deformation can be obtained by Taylor expansion of Eq. \eqref{eq:trajectory1},
\begin{equation}
  x_\text{max} = v^{4/5}\sum_{k=0}^\infty b_k\beta^kv^{k/5}
  \label{eq:xmax}
\end{equation}
with the coefficients $b_k$. We will not need these coefficients explicitly, they can, nevertheless, be found in table \ref{tab:coefficients}. 

\section{Final Velocity For the Na\"ive Condition}

Let us compute the final (na\"{\i}ve) velocity, assuming the end of the collision according to Eq. \eqref{eq:oldbc}. At first glance one might be tempted to compute the duration of collision with an Ansatz like $\tau_c=\tau_c^0 + \delta\tau_c$ and solve for the correction terms by performing a Taylor expansion around the undamped duration of collision. This method, however, fails due to the aforementioned singularity at $x=0$. Instead we compute the final velocity indirectly: as we have an expression which is definitely valid for the first part up to the maximal compression we can construct the full solution by a kind of backward-shooting method. We start at the end of the collision where $\dot{x}=-v^\prime$ (the final velocity $v^\prime$ being unknown yet) and let the time run backwards. The equation of motion for this inverse collision is identical to Eq. \eqref{eq:trajectory}
\begin{equation}
  \label{eq:eqnofmotioninv}
  \begin{split}
  &\ddot{x}_\text{inv} - \beta v^{\prime{\,-1/5}}\dot{x}_\text{inv}\sqrt{x_\text{inv}} + v^{\prime\,{-2/5}}x_\text{inv}^{3/2} = 0\\
  &x_\text{inv}(0) = 0\,;~~~\dot{x}_\text{inv}(0) = v^{\prime\,{4/5}}\,,
  \end{split}
\end{equation}
except for the sign of the damping parameter $\beta$, since the inverse collision (in inverse time) is an accelerated collision. 
Consequently, the trajectory of the inverse problem can be obtained from the solution of the direct collision, Eq. \eqref{eq:trajectory2},  by simply substituting $\beta\to-\beta$ and $v\to v^\prime$.
\begin{multline}
  x_\text{inv} = v^{\prime\,{4/5}}\left[\tau+\frac{4}{15}\beta v^{\prime\,{1/5}}\tau^{5/2} - \frac{4}{35}\tau^{7/2}\right.\\
  \left. + \frac{1}{15}\beta^2 v^{\prime\,{2/5}}\tau^4 - \frac{3}{70}\beta v^{\prime\,{1/5}}\tau^5\right.\\
  \left. + \frac{38}{2475}\beta^3 v^{\prime\,{3/5}}\tau^{11/2}\right] + {\cal O}\left(\tau^6\right)
  \label{eq:xinvraw}
\end{multline}
The same is true for the maximal compression of the inverse collision,
\begin{equation}
  x_\text{max}^\text{inv} = v^{\prime\,{4/5}}\sum_{k=0}^\infty (-1)^kb_k\beta^k v^{\prime\,{k/5}}\,,
\end{equation}
with the same numerical coefficients $b_k$ as in Eq. \eqref{eq:xmax}. 

As the inverse collision problem is just a reformulation for the original collision problem both maximal compressions have to be the same,
\begin{equation}
  x_\text{max} = x_\text{max}^\text{inv}\,,
  \label{eq:equalxmax}
\end{equation}
which is an equation for $v^\prime$. 
From these arguments the choice of our length scale, Eq. \eqref{eq:scales}, becomes evident: choosing the natural unit of length, $k^{-2/5} v^{-1/5}$, the direct and the inverse collision problem would have different length scales as the initial velocity of the inverse collision is $v^\prime \ne v$. 

In order to solve Eq. \eqref{eq:equalxmax} for $v^\prime$ we use the Ansatz
\begin{equation}
  v^\prime = v + \beta \delta v_1 + \beta^2 \delta v_2 + \ldots
\end{equation}
and solve for the corrections $\delta v_i$. Using the definition Eq. \eqref{eq:eps1} this yields the coefficient of restitution of the form 
\begin{equation}
  \varepsilon(v)^\text{na\"{\i}ve} = 1 + c_1\beta v^{1/5} + c_2\beta^2v^{2/5} + \ldots
  \label{eq:epsofvnaive}
\end{equation}
Note that we determined the final velocity $v^\prime$ at $x=0$, that is, this result for $\varepsilon(v)$ corresponds to the condition Eq. \eqref{eq:oldbc} for the end of the collision. Based on the trajectory derived so far, in the next section we will derive the coefficient of restitution that corresponds to Eq. \eqref{eq:newbc}.

The calculation of the coefficients $c_k$ in Eq. \eqref{eq:epsofvnaive} is explained in Appendix \ref{sec:appendix:maxcomp}, the numerical values of the first coefficients are shown in Table \ref{tab:coefficients}.  

\section{Premature end of the collision}

Up to here we calculated the solution of the equation of motion, Eq. \eqref{eq:viscoNewton}, in the interval $(\xi=0, \dot{\xi}=v)$ (start of the collision) to $(\xi=0, \dot{\xi}=v^\prime)$ (end of the collision) or the scaled Equation \eqref{eq:trajectory} in the corresponding interval $x=0$ in the beginning and $x=0$ in the end, respectively. The velocity at the end of this trajectory, $v^\prime$, led us to the coefficient of restitution corresponding to the condition Eq. \eqref{eq:oldbc}.

As discussed before, however, the velocity $v^\prime$ corresponds to a negative interaction force, in contradiction to the purely repulsive interaction of viscoelastic granular particles.
Therefore, the collision does not end at $x=0$ but before, when the interaction force becomes zero. This condition corresponds to the condition Eq. \eqref{eq:newbc}.  

\begin{figure}[htbp]
  \centering
  \includegraphics[width=0.9\columnwidth,bb=13 9 720 546,clip]{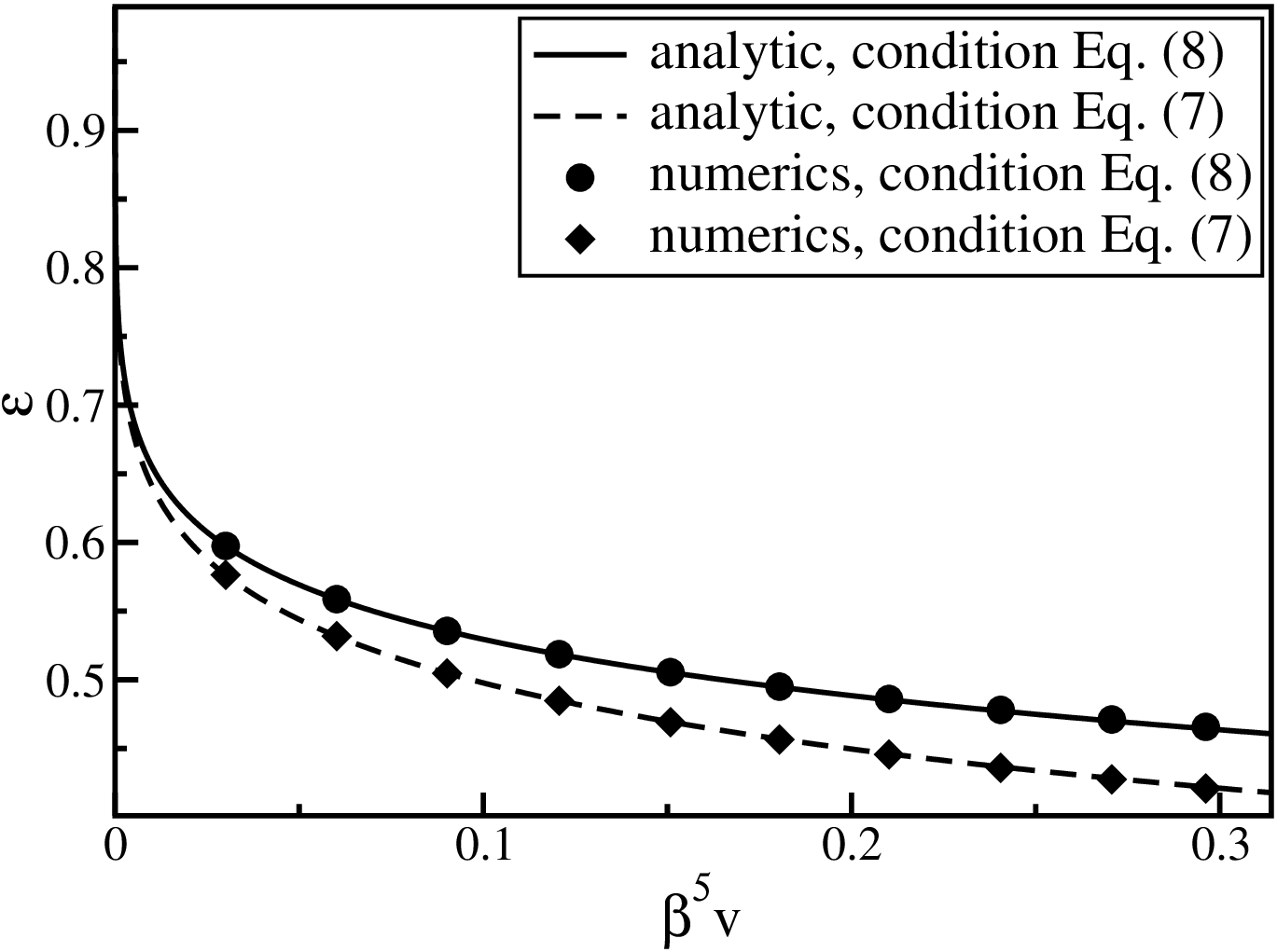}  
  \includegraphics[width=0.9\columnwidth,bb=13 9 720 546,clip]{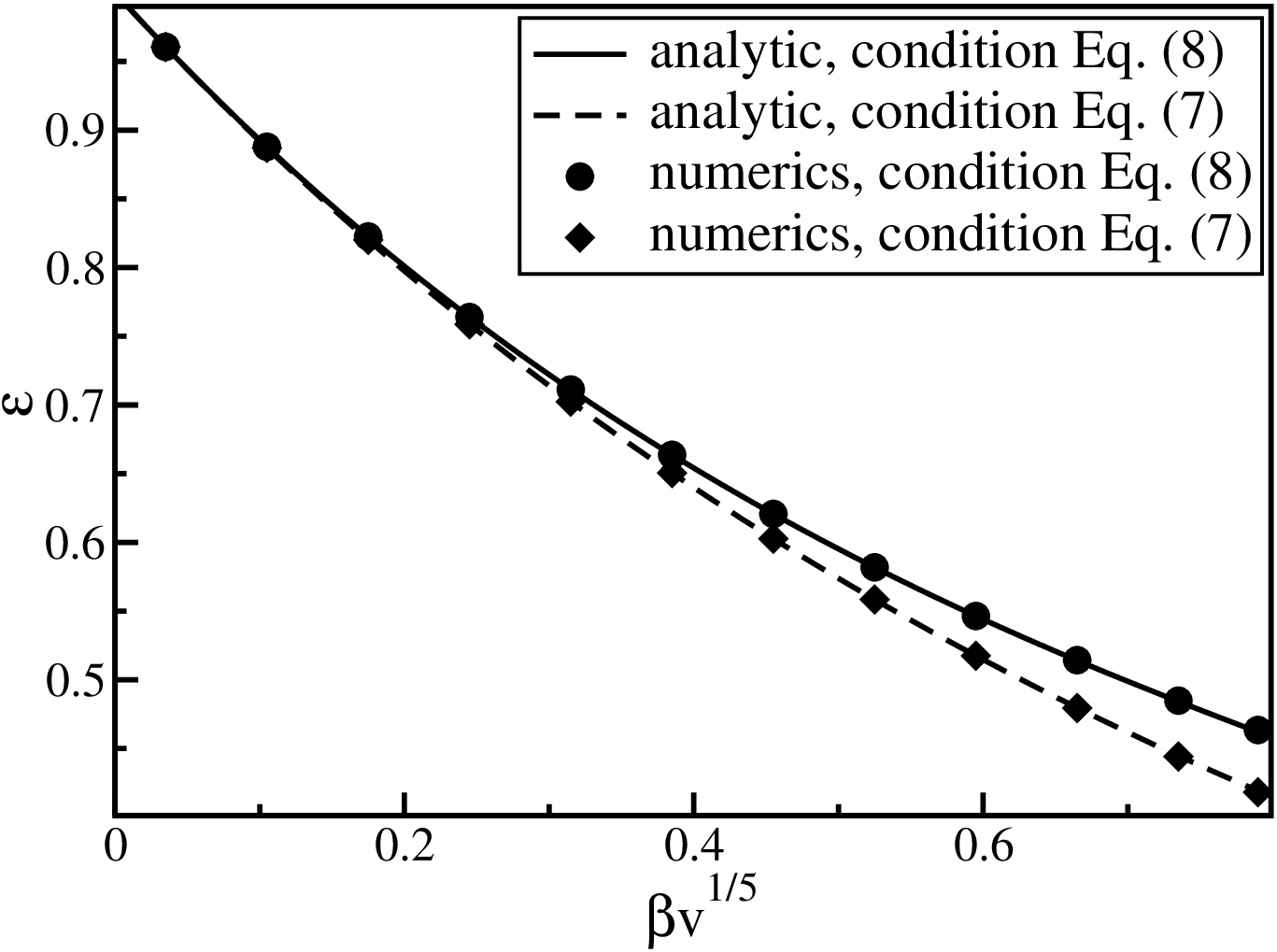}
  \caption{The velocity dependence $\varepsilon(v)$ for both conditions for the end of the collision, Eq. \eqref{eq:oldbc} (na\"ive) and Eq. \eqref{eq:oldbc} (improved). The upper panel shows the dependence on velocity, the lower panel shows the dependence on $\beta v^{1/5}$. For both panels the interval shown in the abscissa is (almost) equivalent. The numerical solution of Newton's equation of motion, Eq. \eqref{eq:trajectory}, agrees almost perfectly with the analytical curves shown here. Beyond the shown interval there are increasing discrepancies between theory and simulation.}
  \label{fig:epsofv}
\end{figure}

We take this premature end of collision into account and, thus, look for the earliest point in time $T$ during the inverse collision when the acceleration vanishes. Setting $\ddot{x}_\text{inv}=0$ in Eq. \eqref{eq:eqnofmotioninv} yields
\begin{equation}
  \beta v^{\prime\,{1/5}}\dot{x}_\text{inv}(T) = x_\text{inv}(T)\,.
\end{equation}
For small $\beta v^{\prime\,{1/5}}$ we obtain $T$ to lowest order by approximating $x_\text{inv}$ by $v^{\prime\,{4/5}}\tau$ which yields
\begin{equation}
  T \approx  \beta v^{\prime\,{1/5}}\,.
\end{equation}
The solution to higher order reads:
\begin{multline}
  T = \beta v^{\prime\,{1/5}} + \frac{4}{35}\beta^{7/2} v^{\prime\,{7/10}} 
+\frac{2}{75}\beta^6 v^{\prime\,{6/5}}\\
+\frac{21271}{2734875} \beta^{17/2} v^{\prime\,{17/10}}
\ldots
\end{multline}
The details of this calculation can be reviewed in Appendix \ref{sec:appendix:abriss}. The value of $\dot{x}_\text{inv}$ at this point in time is
\begin{equation}
  \dot{x}_\text{inv}(T) = v^{\prime\,{4/5}}\left[1+\frac{4}{15}\beta^{5/2} v^{\prime\,{1/2}}
+ \frac{11}{210}\beta^5v^\prime + \ldots\right]
\end{equation}
Going back to the original units of time we obtain the final velocity for the case of the condition Eq. \eqref{eq:newbc}, 
\begin{equation}
  \begin{split}
  v^{\prime\prime} =& \dot{\xi}_\text{final}\\
  =& v^\prime\left[1+\frac{4}{15}\beta^{5/2} v^{\prime\,{1/2}} + \frac{11}{210}\beta^5v^\prime + \ldots\right]
\end{split}
\end{equation}
Inserting the expression for $v^\prime$ one arrives at the final solution
\begin{eqnarray}
  \varepsilon &=& 1 - 1.153\beta v^{1/5} + 0.798\beta^2v^{2/5} + 0.267\beta^{5/2}v^{1/2} + \ldots\nonumber\\
  &=& 1 + \sum_{k=0}^\infty h_k\beta^{k/2}v^{k/10}
  \label{eq:epsofvcorrect}
\end{eqnarray}
The details of this computation are shown in Appendix \ref{sec:appendix:abriss}. The coefficients $h_k$ are pure numbers; the first 20 of them can be found (to a higher precision as in the expression above) in table \ref{tab:coefficientsgh}. As the coefficient of restitution $\varepsilon$ only depends on $\beta v^{1/5}$ (including half powers of this term) we show the velocity dependence in this universal form in Fig. \ref{fig:epsofv}. 

The analytical results, Eqs. \eqref{eq:epsofvnaive} and \eqref{eq:epsofvcorrect}, are compared with the numerical solution of the equation of motion \eqref{eq:trajectory}. In the interval shown in Fig. \ref{fig:epsofv} the analytical results agree with the numerical results almost perfectly. Beyond the shown interval the solutions start to deviate. 
As an example in physical units we consider a sphere that collides with $\varepsilon=0.8$ at $v=1\,$m/sec, e.g. a rubber sphere. By numerically solving Eq. \eqref{eq:epsofvcorrect} we obtain $\beta=0.2\,\text{sec}^{1/5}/\text{m}^{1/5}$. Consequently, the range of velocity shown in Fig. \ref{fig:epsofv}, $\beta v^{1/5}\le0.3$ corresponds to $v\le 7.5$\,m/sec. From the good agreement between the analytical and numerical solutions in this interval we conclude that the range of validity of the solution, Eq. \eqref{eq:epsofvcorrect}, is at least $v\le 7.5$\,m/sec. For materials with smaller damping constant $\beta$ the range of validity is larger. 

Albeit in Fig.  \ref{fig:epsofv} numerical and analytical results almost coincide we note that the deviation for the improved condition, Eq. \eqref{eq:newbc}, exceeds the deviation for the na\"{\i}ve condition by several orders of magnitude. This can be seen from the coefficients $h_k$ (see table \ref{tab:coefficientsgh}) which decrease only slowly for increasing $k$. Thus, to obtain a good precision for $\beta v^{1/5}$ close to unity requires a very large number of coefficients $h_k$. 

\begin{figure}[htbp]
  \centering
  \includegraphics[width=0.9\columnwidth,clip]{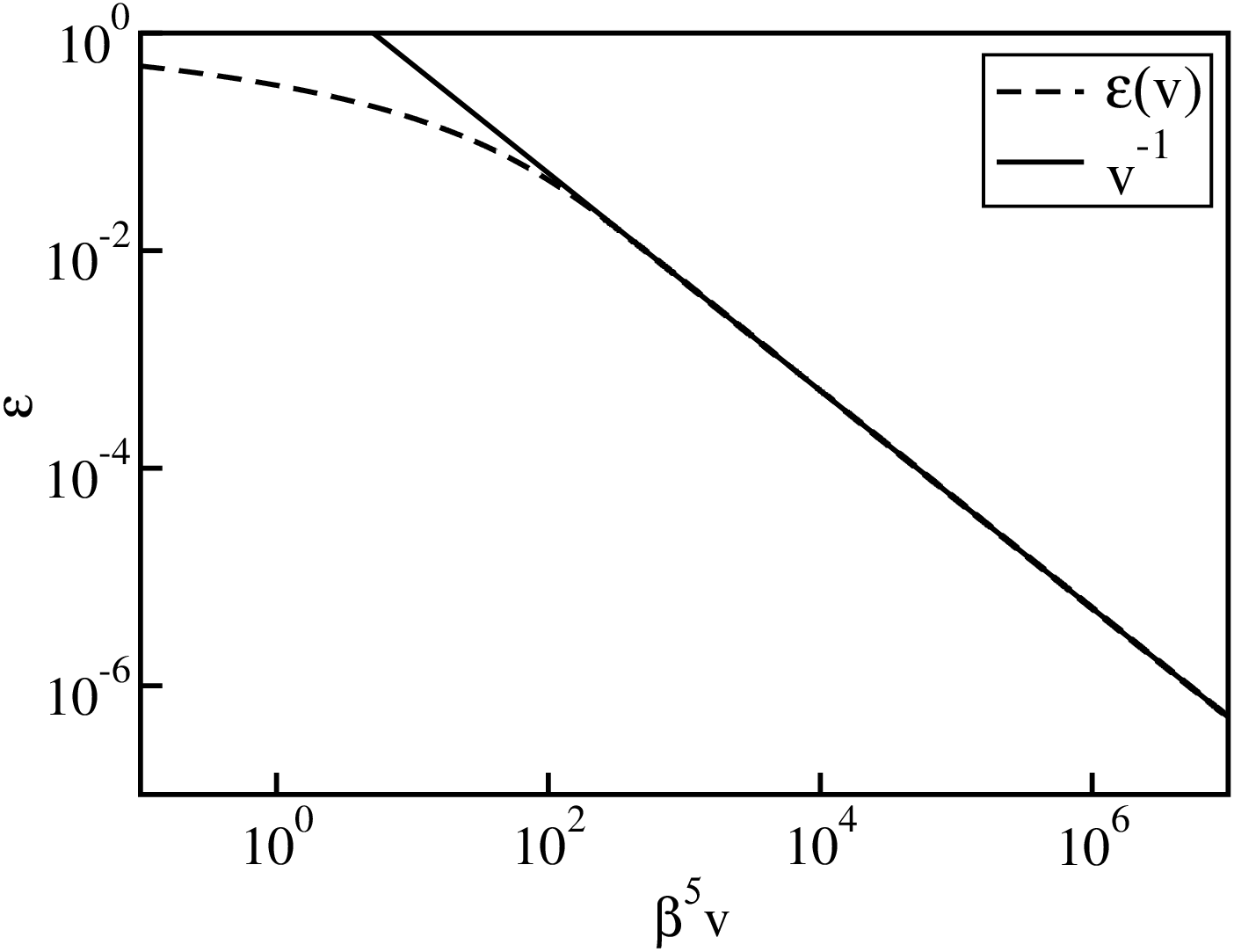}
  \includegraphics[width=0.9\columnwidth,clip]{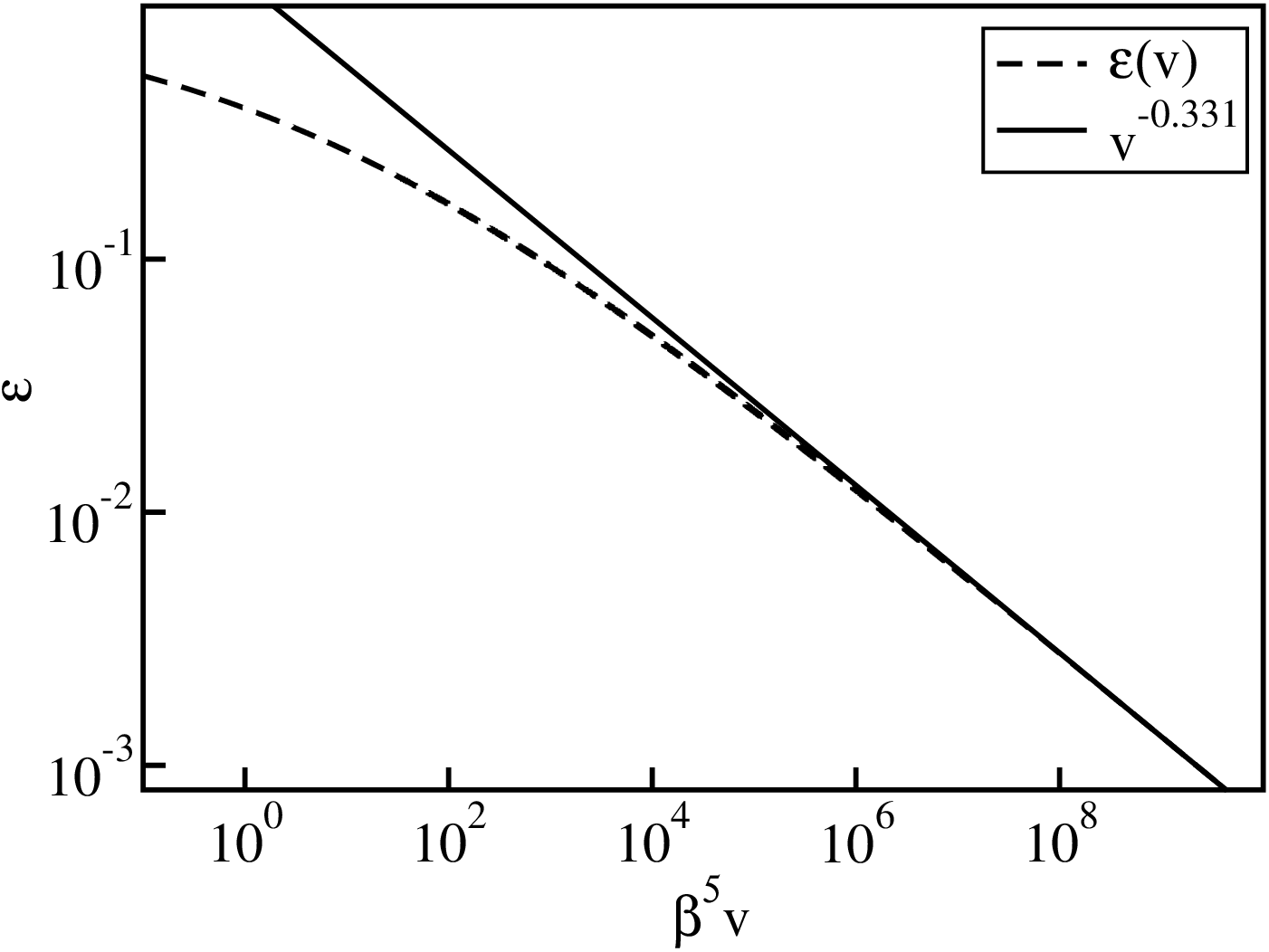}
  \caption{Asymptotics for $\varepsilon(v)$ for the na\"{\i}ve end-of-collision condition, Eq. \eqref{eq:oldbc}, (top) and the improved condition, Eq. \eqref{eq:newbc},  (bottom). Both are representable by a simple power law. For the na\"{\i}ve condition the exponent is $-1$, for the correct condition it is close to $-1/3$.}
  \label{fig:asymptotics}
\end{figure}

For large velocities or large damping both velocity dependencies, Eqs. \eqref{eq:epsofvnaive} and \eqref{eq:epsofvcorrect}, reveal a remarkable difference: For the na\"ive condition, Eq. \eqref{eq:oldbc}, the coefficient of restitution decays asymptotically as $\varepsilon\sim v^{-1}$. For the improved condition, Eq. \eqref{eq:newbc}, the asymptotics is compatible with a power law of $\varepsilon\sim v^{-0.331}$. Both asymptotics are shown in Fig. \ref{fig:asymptotics}.

\section{Conclusion}

We described the collision of a pair of particles which interact repulsively according  to the force law, Eq. \eqref{eq:ViscoForce}, valid for viscoelastic spheres. 
 In a physically consistent description the end of the collision is determined by the instant during the expansion when the interaction force vanishes, $\ddot{\xi}(t)=0$, (a) but not by the na\"{\i}ve condition $\xi(t)=0$ (b) which corresponds to the instant when the distance of the centers of the particles coincides with the sum of their radii. This becomes obvious when looking at the interaction force at the end of the collision: For condition (b) the interaction force becomes attractive which contradicts the precondition of purely repulsive interaction.  The reason for this behavior is the {\em delayed recovery} of the particles, that is, the surfaces of the particles lose contact already slightly before the compressed particles recovered their spherical shape. 

The choice of the condition for the end of the collision, (a) or (b), has a drastic effect on the resulting velocity dependence $\varepsilon(v)$ of the coefficient of normal restitution. Instead of a series in $v^{1/5}$ obtained for the na\"{\i}ve condition (b) \cite{SchwagerPoeschel:1998,RamirezEtAl:1999}, for the physically consistent end-of-collision condition (a) we  obtain a series in $v^{1/10}$ where the odd powers of $v^{1/10}$ are solely due to the end-of-collision rule. The analytical results agree almost perfectly with the numerical integration of Newton's equation of motion for colliding viscoelastic spheres. 

We evaluated the result for $\varepsilon(v)$ for realistic material properties for the cases (a) and (b) and obtained a noticeable difference of up to about 20\%, depending on the material properties. The range of validity of our result was estimated by about 10 m/sec for a soft, rather dissipative material such as rubber. For a more elastic material, corresponding to a larger coefficient of restitution, the range of validity is significantly larger. Our analytical results deviate from the numerical results for $\beta v^{1/5}\gtrsim 0.9$ which may be attributed to the properties of the series, Eq. \eqref{eq:epsofvcorrect}, which converges slowly for large $\beta v^{1/5}$ and whose convergence is not even clear for $\beta v^{1/5}\ge 1$. 

For large impact velocity we can, however, still obtain numerical results which reveal another drastic difference between the conditions (a) and (b): For both conditions, asymptotically  $\varepsilon(v)$ follows a power law. For the na\"{\i}ve condition (b), however, we obtain $\varepsilon\sim v^{-1}$ whereas for the physically consistent condition (a) we find $\varepsilon\sim v^{-1/3}$.

The influence of the end-of-collision condition on the coefficient of restitution for viscoelastic particles is in marked contrast to the corresponding result obtained for the linear dash-pot model \cite{SchwagerPoeschel:2007}. Here the choice of the condition (a) or (b) would result a modified coefficient of restitution which is, nevertheless, independent of the impact velocity in both cases.

\acknowledgments{
This research was supported by a Grant from the G.I.F., the German-Israeli Foundation for Scientific Research and Development.}

\appendix
\section{Computation of the Trajectory}
\label{sec:appendix:trajectory}

Equation \eqref{eq:trajectory} for the trajectory $x(\tau)$ of the particles' relative motion in the scaled variables $x$ and $\tau$,
\begin{equation} 
  \label{eq:App.trajectory}
  \begin{split}
  &\ddot{x} + \beta v^{-1/5}\dot{x}\sqrt{x} + v^{-2/5}x^{3/2} = 0\\
  &x(0)= 0\,;~~~\dot{x}(0) = v^{4/5}
  \end{split}
\end{equation}
is solved by series expansion. As explained in the text, an expansion in powers of $\tau$ fails, instead we expand in powers of $\sqrt{\tau}$. Using the Ansatz
\begin{equation}
\label{eq:App.Ransatz}
  x(\tau) = v^{4/5}\tau \left[1+R(\tau)\right]
\end{equation}
Eq. \eqref{eq:App.trajectory} turns into
\begin{equation}
  \begin{split}
&  2\dot{R} + \tau\ddot{R}+ \beta v^{1/5}\tau^{1/2}\left(1+R + \tau\dot{R}\right)\sqrt{1+R}\\
&  ~~~~~~~+ \tau^{3/2}\left[1+R\right]^{3/2} = 0\\
&    R(0)=0\,;~~~\dot{R}(0)=0\,.
  \end{split}
  \label{eq:App.eqnofmotioninR}
\end{equation}
The term $R$ may be expanded in powers of $\sqrt{\tau}$,
\begin{equation}
  \label{eq:App.Rexpansion}
  R(\tau)=a_0+a_1\tau^{1/2}+a_2\tau^{1}+a_3\tau^{3/2}\ldots
\end{equation}
Inserting Eq. \eqref{eq:App.Rexpansion} into Eq. \eqref{eq:App.trajectory} and comparing equal powers of $\tau^{1/2}$ we find $a_0=a_1=a_2=0$, that is, the first non-trivial contribution is ${\cal O}(\tau^{3/2})$. This fact simplifies the subsequent computer algebra considerably.

We determine the coefficients $a_3$, $a_4, \ldots$ in escalating order using an iterative procedure. In the first step we determine $a_3$ while $a_i$ ($i>3)$ stay undetermined. The corresponding term for $R$ of the order 3 is denominated by $R_3\equiv a_3\tau^{3/2}+{\cal O}(\tau^2)$ the next order is $4$ with $R_4\equiv a_4\tau^2+{\cal O}(\tau^{5/2})$, etc. In other words, $R_i$ contains all contributions of order ${\cal O}(\tau^{i/2})$ and higher. In each step $i$ of the iteration we derive a differential equation $G_i[R_i,\dot{R}_i,\ddot{R}_i]=0$ for $R_i$. 

We demonstrate the procedure for the first terms of a series up to the term $a_9 \tau^{9/2}$. For the first step, $i=3$, we expand $(1+R)^{1/2}$ and $(1+R)^{3/2}$ in Eq. \eqref{eq:App.eqnofmotioninR} up to the necessary order for $R_3$. Since $N=9$ and the lowest order of $\tau$ in $R$ is 3, we need the expansion up to the third term,
\begin{equation}
  \label{eq:App.expansion}
  \begin{split}
    \sqrt{1+R_3} =& 1 + \frac{R_3}{2}-\frac{R_3^2}{8} + \frac{R_3^3}{16}\\
    (1+R_3)^{3/2} =& 1 + \frac{3R_3}{2} + \frac{3R_3^2}{8} - \frac{R_3^3}{16}
  \end{split}
\end{equation}
Equation \eqref{eq:App.eqnofmotioninR} reads then
\begin{multline}
G_3\left[R_3\right] \equiv 2\dot{R}_3 + \tau\ddot{R}_3 + \beta v^{1/5}\tau^{1/2} \times\\
\times \left(1+\frac32R_3 + \frac38 R_3^2 + \tau\dot{R}_3 +\frac12\tau\dot{R}_3 R_3 \right)\\ 
+\tau^{3/2} \left(1 + \frac32R_3 +  \frac38 R_3^2\right) = 0\,,
\label{eq:App.G3}
\end{multline}
where terms of order $\tau^{10/2}$ and higher are neglected. 

The desired coefficient $a_3$ is now isolated by the formal transformation
\begin{equation}
  R_3 = a_3\tau^{3/2} + R_4
  \label{eq:App.iteration3}
\end{equation}
which establishes the first iteration step. In general, we replace
\begin{equation}
  R_i = a_i\tau^{i/2} + R_{i+1}\,,
\end{equation}
insert this into $G_i(R_i)=0$ where only terms of relevant order are taken into account. Then we consider the term ${\cal O}\left(\tau^{i/2-1}\right)$ and determine $a_i$. After substituting $a_i$ back into $G_i$ we are left with the next order equation  $G_{i+1}\left(R_{i+1}\right)=0$ which is then solved in the same way, etc.

We insert Eq. \eqref{eq:App.iteration3} into Eq. \eqref{eq:App.G3} and obtain
\begin{widetext}
\begin{multline}
  \label{eq:App.iter3}
  2\dot{R}_4+\tau\ddot{R}_4+
  \left(\frac{15}{4}a_3 + \beta v^{1/5} \right)\tau^{1/2} 
  +\tau^{3/2}
  +3 \beta v^{1/5}a_3\tau^2
  +\frac{3}{2}a_3\tau^{3}
  +\frac{9}{8}\beta v^{1/5} a_3^2 \tau^{7/2}
  +\frac{3}{8}a_3^2\tau^{9/2}\\
  +\frac{3}{2}\beta v^{1/5}R_4\tau^{1/2}
  +\frac{3}{2}R_4\tau^{3/2} + \beta v^{1/5} \dot{R}_4 \tau^{3/2}
  +\frac{3}{2}\beta v^{1/5} a_3 R_4 \tau^2
  +\frac{3}{8}\beta v^{1/5} R_4^2 \tau^{1/2} =0
\end{multline}
where again terms of irrelevant order were skipped. The terms in brackets of lowest order 1/2 ($i/2-1$ in general) allows for the computation of the first non-trivial coefficient $a_3=-(4/15)\beta v^{1/5}$\,.
We insert $a_3$ into Eq. \eqref{eq:App.iter3} to obtain the next equation for the computation of $a_4$:
\begin{multline}
  \label{eq:App.G4}
G_4 \left[R_4\right]\equiv   
2 \dot{R}_4+\tau \ddot{R}_4
+\frac{3}{2}\beta v^{1/5} R_4 \tau^{1/2} + \frac{3}{8}\beta v^{1/5} R_4^2 \tau^{1/2} 
+\tau^{3/2} + \frac{3}{2} R_4 \tau^{3/2} +\beta v^{1/5} \dot{R}_4 \tau^{3/2} 
-\frac{2}{5}\beta^2 v^{2/5} R_4 \tau^2 \\
-\frac{4}{5}\beta^2 v^{2/5} \tau^2 
-\frac{2}{5} \beta v^{1/5} \tau^3 
+\frac{2}{25} \beta^3 v^{3/5} \tau^{7/2}
+\frac{2}{75} \beta^2 v^{2/5} \tau^{9/2}=0
\end{multline}
The next iteration step  $R_4=a_4\tau^2 + R_5$
leads to
\begin{multline}
  \label{eq:App.iter4}
2\dot{R}_5 + \tau \ddot{R}_5 +  6 a_4\tau + 
  \tau^{3/2} - \frac{4}{5}\beta^2 v^{2/5} \tau^{2} + \frac{7}{2}\beta v^{1/5} a_4 \tau^{5/2} 
  - \frac{2}{5} \beta v^{1/5} \tau^3 
  + \left(\frac{3}{2}a_4+\frac{2}{25}\beta^3 v^{3/5}\right)\tau^{7/2}\\
+\frac{3}{2}\beta v^{1/5}R_5\tau^{1/2} + \frac{3}{2}R_5 \tau^{3/2} + \beta v^{1/5}\dot{R}_5 \tau^{3/2} 
-\frac{2}{5}\beta^2 v^{2/5} R_5 \tau^2 
-\frac{2}{5}\beta^2 v^{2/5} a_4 \tau^4
+\frac{3}{8} \beta v^{1/5} a_4^2 \tau^{9/2} + \frac{2}{75} \beta^2 v^{2/5} \tau^{9/2}
=0\,.
\end{multline}
From the terms of lowest order we find $0=6a_4\tau$, that is $a_4 = 0$.
We insert this into Eq. \eqref{eq:App.iter4}:
\begin{multline}
  \label{eq:App.G5}
  G_5\left[R_5\right] \equiv 2\dot{R}_5 + \tau \ddot{R}_5 + \tau^{3/2} - \frac{4}{5}\beta^2v^{2/5}\tau^2 - \frac{2}{5}\beta v^{1/5}\tau^3 + \frac{2}{25}\beta^3 v^{3/5}\tau^{7/2} + \frac{3}{2}\beta v^{1/5}R_5\tau^{1/2} + \frac{3}{2}R_5\tau^{3/2}\\ + \beta v^{1/5} \dot{R}_5 \tau^{3/2} 
-\frac{2}{5}\beta^2 v^{2/5} R_5 \tau^2
+\frac{2}{75}\beta^2 v^{2/5} \tau^{9/2}  = 0\,.
\end{multline}
With $R_5= a_5 \tau^{5/2}+R_6$
the latter equation turns into
\begin{multline}
2\dot{R}_6+ \tau \ddot{R}_6 
+\left(\frac{35}{4}a_5 + 1\right)\tau^{3/2} 
- \frac{4}{5}\beta^2 v^{2/5} \tau^2 + \left(4\beta v^{1/5} a_5 - \frac{2}{5}\beta v^{1/5}\right) \tau^3 
+\frac{2}{25} \beta^3 v^{3/5} \tau^{7/2}\\
+\frac{3}{2}a_5\tau^4 
+\left(
-\frac{2}{5}\beta^2v^{2/5}a_5
+\frac{2}{75}\beta^2 v^{2/5} \right) \tau^{9/2} 
+ \frac{3}{2} \beta v^{1/5} R_6\tau^{1/2} 
+\beta v^{1/5}\dot{R}_6\tau^{3/2} 
+\frac{3}{2}R_6\tau^{3/2} 
= 0\,.
\end{multline}
From the lowest order terms ${\cal O}\left(\tau^{3/2}\right)$ we obtain $a_5= -4/35$. We insert
\begin{multline}
  \label{eq:App.G6}
  G_6\left[R_6\right]\equiv 2 \dot{R}_6 
  + \tau \ddot{R}_6 
  - \frac{4}{5}\beta^2 v^{2/5}\tau^2 
  - \frac{6}{7}\beta v^{1/5}\tau^3 
  + \frac{2}{25}\beta^3 v^{3/5} \tau^{7/2} 
  - \frac{6}{35}\tau^4 
  + \frac{38}{525}\beta^2 v^{2/5}\tau^{9/2}
  + \frac{3}{2}\beta v^{1/5} R_6\tau^{1/2} \\
  + \frac{3}{2}R_6\tau^{3/2} 
  +\beta v^{1/5}\dot{R}_6 \tau^{3/2}
= 0
\end{multline}
iterate, $R_6 = a_6\tau^3 + R_7$, and obtain
\begin{multline}
  \label{eq:App.R6insert}
2\dot{R}_7 + \ddot{R}_7\tau 
+ \left(12 a_6 - \frac{4}{5}\beta^2 v^{2/5}\right) \tau^2 
- \frac{6}{7}\beta v^{1/5} \tau^3 
+ \left(\frac{9}{2} \beta v^{1/5} a_6 + \frac{2}{25}\beta^3 v^{3/5}\right) \tau^{7/2} 
- \frac{6}{35}\tau^4 \\
+ \left(\frac{3}{2}a_6 + \frac{38}{525}\beta^2 v^{2/5}\right) \tau^{9/2}
+ \frac{3}{2}\beta v^{1/5} R_7 \tau^{1/2}  
= 0\,.
\end{multline}
From the term of lowest order we find  $a_6 = \beta^2 v^{2/5}/15$. 
We insert $a_6$ into Eq. \eqref{eq:App.R6insert} for the next order equation
\begin{equation}
  \label{eq:G7}
  G_7\left[R_7\right]\equiv 2 \dot{R}_7 + \tau \ddot{R}_7 
  - \frac{6}{7} \beta v^{1/5}\tau^3 
  + \frac{19}{50} \beta^3 v^{3/5} \tau^{7/2} 
  -\frac{6}{35}\tau^4 
  + \frac{181}{1050}\beta^2 v^{2/5} \tau^{9/2}
  + \frac{3}{2} \beta v^{1/5} R_7 \tau^{1/2}
= 0\,.
\end{equation}
Iterating  $R_7= a_7 \tau^{7/2} + R_8$ yields
\begin{equation}
  \label{eq:App.R7insert}
  2 \dot{R}_8 + \ddot{R}_8\tau 
  +\frac{63}{4} a_7 \tau^{5/2} 
  - \frac{6}{7}\beta v^{1/5}\tau^3 
  + \frac{19}{50}\beta^3 v^{3/5} \tau^{7/2} 
  + \frac{3}{2}\beta v^{1/5} a_7\tau^4 
  -\frac{6}{35}\tau^4 
  + \frac{181}{1050}\beta^2 v^{2/5}\tau^{9/2} 
  + \frac{3}{2}\beta v^{1/5}R_8\tau^{1/2}
= 0
\end{equation}
and from lowest order $\sim \tau^{5/2}$ we obtain: $a_7=0$.
We insert this solution, and substitute $R_8 = a_8 \tau^4 + R_9$:
\begin{equation}
  \label{eq:App.R9insert}
2 \dot{R}_9 + \tau \ddot{R}_9 
+ \left(20 a_8 -\frac{6}{7}\beta v^{1/5}\right) \tau^3
+\frac{19}{50}\beta^3 v^{3/5} \tau^{7/2}
-\frac{6}{35} \tau^4
+ \left(\frac{3}{2}\beta v^{1/5} a_8 
+\frac{181}{1050}\beta^2 v^{2/5} \right) \tau^{9/2}=0
\end{equation}
\end{widetext}
We insert the solution $a_8 = \frac{3}{70} \beta v^{1/5}$:
\begin{multline}
  G_9\left[R_9\right] \equiv  2\dot{R}_9 + \tau \ddot{R}_9 
+\frac{19}{50} \beta^3 v^{3/5} \tau^{7/2}
-\frac{6}{35}\tau^4\\
+\frac{71}{300}\beta^2 v^{2/5} \tau^{9/2} =0
\end{multline}
replace $R_9 = a_9 \tau^{9/2} + R_{10}$,
\begin{multline}
  2\dot{R}_{10} + \ddot{R}_{10} \tau 
  +\left(\frac{99}{4} a_9 + \frac{19}{50} \beta^3 v^{3/5}\right) \tau^{7/2} \\
  - \frac{6}{35} \tau^4 
  + \frac{71}{300} \beta^2 v^{2/5} \tau^{9/2}=0
\end{multline}
and obtain $a_9=-(38/2475)\beta^3 v^{3/2}$. Inserting this solution and substituting 
$R_{10}=a_{10} \tau^5 + R_{11}$ yields
\begin{equation}
  2 \dot{R}_{11} +\tau \ddot{R}_{11} 
  + \left(30 a_{10} -\frac{6}{35}\right) \tau^4
  + \frac{71}{300} \beta^2 v^{2/5} \tau^{9/2}
  =0
\end{equation}
and, thus, $a_{10}=1/175$ which is the last coefficient which can be obtained from the expansion, Eq. \eqref{eq:App.expansion} 

To achieve an acceptable accuracy of the final result, the expansion Eq. \eqref{eq:epsofvcorrect} has to be performed up to high orders in $\beta v^{1/5}$. To accurately compute the necessary coefficients $h_k$ one needs accurate functions $x_k$ of the same index $k$. For the chosen accuracy (20 coefficients $h_k$) the expansion of the trajectory has to be performed up to an order as large as 150. We employ computer algebra (maple) which turns the described
algorithm into only a few lines of code. For the computation we
abbreviate $A \equiv \beta v^{1/5}$, $s\equiv \sqrt{\tau}$, \verb|Rd|
and \verb|Rdd| stand for $dR/d\tau$ and $d^2R/d\tau^2$, and \verb|N|
is the order of the expansion.
\begin{verbatim}
restart;
N := 150;
dgl:=2*Rd+s^2*Rdd+(A*s+s^3)*(1+R)^(3/2)
     +A*s^3*Rd*sqrt(1+R);
dgl:=convert(taylor(dgl,R,N),polynom):
solution:=0;
for i from 3 to N do
  dgl:=subs(Rdd=(i*(i-2)/4)*a*s^(i-4)+Rdd, 
    Rd=(i/2)*a*s^(i-2)+Rd,R=a*s^i+R,dgl):
  dgl:=mtaylor(dgl,[R,s,Rd,Rdd],N,[i,1,i,i]):
  tmp:=expand(coeff(coeff(coeff(dgl,R,0),
    Rdd,0),Rd,0));
  asol:=solve(coeff(tmp,s,i-2),a):
  print(i, asol):
  dgl:=simplify(subs(a=asol,dgl)):
  solution:=solution+asol*s^i:
end do:
solution:=v^(4/5)*s^2*(1+solution):
solution:=subs(A=beta*v^(1/5),solution):
fout:=fopen("./solution",WRITE);
fprintf(fout,"%a\n",solution);
fclose(fout);
\end{verbatim}
%The code and a table of the higher-order coefficients $a_i$ 
%are available online \cite{online}

\section{Time and Value of Maximal Compression and the series $\varepsilon_\text{naive}(v)$}
\label{sec:appendix:maxcomp}

The first ingredient for the actual computation of $\varepsilon_\text{naive}(v)$ is the maximum compression. To this end, we first compute at which time $\tau_\text{max}$ this maximum compression is achieved. 
The time of maximum compression will be determined by Taylor-expansion of the expression
\begin{equation}
  \dot{x}\left(\tau_\text{max}^{0} + \delta \tau\right) = 0\,. 
\end{equation}
Here the time $\tau_\text{max}^0$ of maximum compression of the undamped collision as given by Eq. \eqref{eq:tchhalf} is taken as a reference. 
In terms of the universal functions $x_i$ as introduced in Eq. \eqref{eq:trajectory1} the Taylor expansion takes the form
\begin{equation}
  \dot{x}\left(\tau_\text{max}^0 + \delta\tau\right) = v^{4/5}\sum_{i=0}^{n_\beta}\beta^iv^{i/5}\sum_{k=0}^{n_\beta}\frac{d^{k+1}x_i}{d\tau^{k+1}}\frac{\delta\tau^k}{k!}\label{eq:dotxTaylor}
\end{equation}
which motivates the representation of $\delta \tau$ as a series of the form
\begin{equation}
  \delta\tau = \sum_{n=1}^{n_\beta}a_n\beta^nv^{n/5}\,.
  \label{eq:defdeltatau}
\end{equation}
We insert Eq. \eqref{eq:defdeltatau} into the Taylor expansion, collect coefficients in powers of $\beta$ and solve successively for $a_n$. The result is shown in Tab. \ref{tab:coefficients}.

In the same way the maximum compression $x_\text{max}$ can be computed by performing the Taylor-expansion of $x(\tau^0_\text{max}+\delta\tau)$ which is of the form
\begin{equation}
  x_\text{max} = v^{4/5}\sum_{i=0}^{n_\beta}\beta^iv^{i/5}\sum_{k=0}^{n_\beta}\frac{d^kx_i}{d\tau^k}\frac{\delta\tau^k}{k!}\label{eq:hmaxTaylor}
\end{equation}
suggesting the series
\begin{equation}
  x_\text{max} = v^{4/5}\sum_{n=0}^{n_b}b_n\beta^nv^{n/5}
\end{equation}
The coefficients $b_n$ are shown as well in Tab. \ref{tab:coefficients}.
The first coefficient $b_0$ is the maximum compression for the undamped problem. To actually compute the coefficient of normal restitution without regard of the premature loss of contact we have to match the maximum compression of the direct and the inverse collision, i.e. we have to solve
\begin{equation}
  x_\text{max}(v) = x_\text{max}^\text{inv}\left(v^\prime\right)
  \label{eq:match}
\end{equation}
for $v^\prime$ with
\begin{eqnarray}
  x_\text{max}(v) &=& v^{4/5}\sum_{n=0}^{n_\beta}b_n\beta^nv^{n/5}\\
  x_\text{max}(v^\prime) &=& v^{\prime\,{4/5}}\sum_{n=0}^{n_\beta}(-1)^nb_n\beta^n v^{\prime\,{n/5}}
\end{eqnarray}
In the Maple program the function $x_\text{max}(v)$ is called \verb|h(v)|, the function $x_\text{max}(v)$ is called \verb|hm(v)|. Using the Ansatz
\begin{equation}
  v^{\prime} = v\left(1+\sum_{n=1}^{n_\beta}c_n\beta^nv^{n/5}\right) = v\varepsilon_\text{naive}(v)
\end{equation}
we can solve for $c_n$ by expanding the expression Eq. \eqref{eq:match} for small $\beta$ and collect orders. The first $c_k$ are shown in  Tab. \ref{tab:coefficients}.
\begin{table}
\caption{The first numerical coefficients of the expansions Eq. \eqref{eq:defdeltatau}}
  \vspace*{0.2cm}
  \newcolumntype{h}[1]{D{.}{.}{#1}}
  \begin{center}
    \begin{tabular}{h{0}@{~}|@{~}h{12}@{~}|@{~}h{12}@{~}|@{~}h{11}}
      i & a_i               & b_i              & c_i           \\
      \hline
      0 &                   & 1.093362074      &               \\  
      1 &  -0.2867471220    & -0.5044548926    & -1.153448854  \\
      2 &  0.1048589922     & 0.2840430192     & 0.7982665553  \\
      3 &  -0.04868406400   & -0.1702207776    & -0.5228825609 \\
      4 &  0.02543116890    & 0.1055007088     & 0.3487426678  \\
      5 &  -0.01423658282   & -0.06684871371   & -0.2330981260 \\
      6 &  0.008337660013   & 0.04303949229    & 0.1566821477  \\
      7 &  -0.005039737366  & -0.02805108430   & -0.1058187828 \\
      8 &  0.003118137108   & 0.01846085121    & 0.07176528242 \\
      9 &  -0.001964027745  & -0.01224618562   & -0.04885717237\\
      10&  0.001254701962   & 0.008177589114   & 0.03337347194 \\
    \end{tabular}
  \end{center}
  \label{tab:coefficients}
\end{table}

\begin{verbatim}
restart; Digits := 20;
nb := 20;
fin := fopen("./solution", READ):
xin := fscanf(fin, "%a"):
x := simplify(subs(s = sqrt(t), xin[1])):
tchalf := (4/5)^(3/5)*GAMMA(2/5)*GAMMA(1/2)/
          (2*GAMMA(9/10)):
x := subs(t = tchalf+dt, x):
xdot := evalf(taylor(diff(x, dt),dt=0,nb)):
xdot := convert(xdot, polynom):
dt := sum(a['i']*beta^'i'*v^((1/5)*'i'), 
          'i' = 1 .. nb):
for i to nb do 
  a[i]:= solve(coeff(xdot, beta, i), a[i]) 
od:
hh := convert(evalf(taylor(x, beta, nb+1)), 
              polynom):
xmax := unapply(hh, v):
xmaxinv := unapply(subs(beta = -beta, hh), v):
u := v*(1+sum(c['k']*beta^'k'*v^((1/5)*'k'), 
              'k' = 1 .. nb)):
d := convert(taylor(xmax(v)-xmaxinv(u), 
                    beta, nb+1), polynom):
for i to nb do 
  c[i]:= solve(coeff(d, beta, i),c[i]) 
od:
fout := fopen("./coefficients", WRITE):
for i to nb do 
  fprintf(fout, "%a\n", c[i]) 
od:
fclose(fout):
\end{verbatim}

\section{Premature loss of contact}
\label{sec:appendix:abriss}

As the moment of actual loss of contact is close to the na\"ive end of contact we will use the inverse collision to compute the time $T$ and velocity $v^{\prime\prime}$ at loss of contact. Using the condition $\ddot{x}_\text{inv}=0$ we obtain the equation for $T$:
\begin{equation}
  \beta v^{\prime\,{1/5}}\dot{x}_\text{inv}(T) = x_\text{inv}(T)
  \label{eq:definitionofT}
\end{equation}
Approximating $x_\text{inv}$ as $v^{\prime\,{4/5}}T$ we obtain the leading order of
\begin{equation}
  T = \beta v^{\prime\,{1/5}}
\end{equation}
After canceling the common prefactor $v^{\prime\,{4/5}}$ Eq. \eqref{eq:definitionofT} does only depend on the combination $\beta v^{\prime\,{1/5}}$. Therefore, one can easily guess the principal form of $T$:
\begin{equation}
  T = \beta v^{\prime\,{1/5}} + \sum_{k=2}^\infty d_k\beta^k v^{\prime\,{k/5}}
\end{equation}
Inserting this Ansatz into Eq. \eqref{eq:definitionofT}, collecting orders and solving for $d_k$ yields
\begin{multline}
  T = \beta v^{\prime\,{1/5}} + \frac{4}{35}\beta^{7/2} v^{\prime\,{7/10}} + \frac{2}{75}\beta^6 v^{\prime\,{6/5}}\\
   + \frac{21271}{2734875}\beta^{17/2} v^{\prime\,{17/10}} + \ldots
\end{multline}
The final solution now reads
\begin{multline}
  v^{\prime\prime} = v\left(1 + \frac{4}{15}\beta^{5/2}v^{1/2} + \frac{13}{150}\beta^5v + \frac{4897}{160875}\beta^{15/2}v^{3/2}\right.\\
  \left. + \frac{453263}{40540500}\beta^{10}v^2 + \ldots\right)
\end{multline}
Inserting the known solution for $v^\prime$ we obtain
\begin{eqnarray}
  v^{\prime\prime} &=& v\left(1+\sum_{k=0}^\infty h_k\beta^{k/2}v^{k/10}\right)\label{eq:hkdef}\\
  \varepsilon(v) &=& 1+\sum_{k=0}^\infty h_k\beta^{k/2}v^{k/10}
\end{eqnarray}
The first values of $h_k$ are tabulated in Table \ref{tab:coefficientsgh}
\begin{table}
\caption{The first numerical coefficients of the expansion \eqref{eq:hkdef}. The coefficients $h_2$ and $h_4$ are identical to the first coefficients in the original expansion $\varepsilon(v)$, i.e. $h_2\equiv c_1$ and $h_4\equiv c_2$.}
  \vspace*{0.2cm}
  \newcolumntype{h}[1]{D{.}{.}{#1}}
  \begin{center}
    \begin{tabular}{h{0}@{~}|@{~}h{12}@{~}}
      i & h_i              \\
      \hline
      0 & 1\\
      1 & 0\\
      2 & -1.153448856\\
      3 & 0\\
      4 & 0.7982665581\\
      5 & 0.2666666667\\
      6 & -0.5228825657\\
      7 & -0.4613795424\\
      8 & 0.3487426737\\
      9 & 0.4523510496\\
      10 & -0.1464314644\\
      11 & -0.3677282992\\
      12 & -0.0432489833\\
      13 & 0.2818042325\\
      14 & 0.1478525872\\
      15 & -0.1794420590\\
      16 & -0.1784660326\\
      17 & 0.06593358882\\
      18 & 0.1713586178\\
      19 & 0.0252498223\\
      20 & -0.1379234986
    \end{tabular}
  \end{center}
  \label{tab:coefficientsgh}
\end{table}

\begin{verbatim}
restart:Order:=20:nd:=4:Digits:=20:
fin:=fopen("./solution",READ):
L:=fscanf(fin,"%a"):fclose(fin):
x:=convert(taylor(L[1],s,Order),polynom):
xinv:=subs(s=sqrt(T),subs(beta=-B*B,x)):
xinvdot:=diff(xinv,T):
eqn:=simplify(xinv-B*B*v^(1/5)*xinvdot):
T:=B^2*v^(1/5)*sum(d['k']*B^(5*'k')*v^('k'/2),
                   'k'=0..nd):
eqn:=expand(eqn):
eqn:=series(eqn,B,2*Order+1):
for i from 0 to nd do 
  d[i]:=solve(coeff(eqn,B,2+5*i),d[i]): 
od:
vpp:=convert(series(v^(1/5)*xinvdot,B,2*Order+1),
             polynom):
fin:=fopen("./coefficients",READ):
for i from 1 to Order do 
  L:=fscanf(fin,"%a"):
  c[i]:=L[1]:
od:
fclose(fin):
vprime:=v*(1+sum(c['k']*B^(2*'k')*v^('k'/5),
                 'k'=1..Order)):
vpp:=convert(series(subs(v=vprime,vpp),
                    B,2*Order+1),
             polynom):
epsilon:=simplify(vpp/v);
fout:=fopen("hk",WRITE):
for i from 1 to 2*Order do 
  h[i]:=simplify(coeff(epsilon,B,i)/v^(i/10)):
  fprintf(fout,"%a,\n",h[i]): 
od:
\end{verbatim}
%\bibliography{EOC}

\end{document}